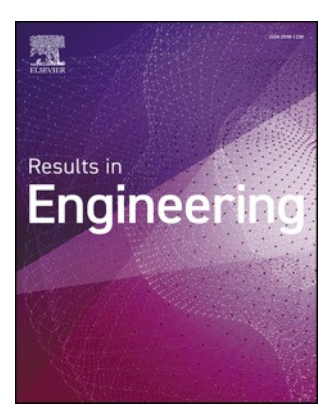

Research paper

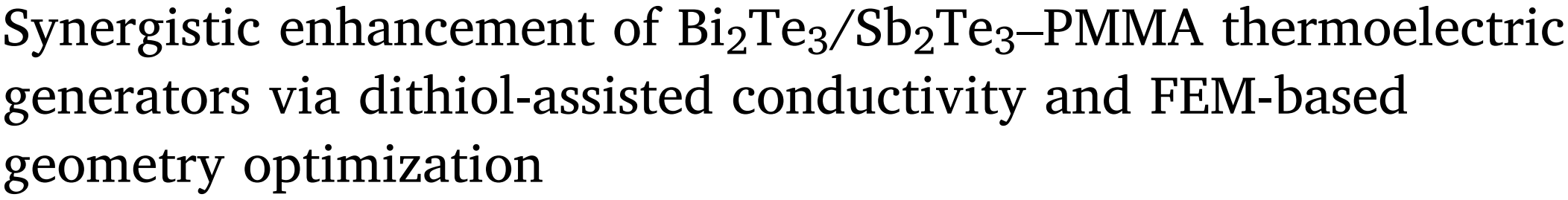

# Synergistic enhancement of $Bi_2Te_3$/$Sb_2Te_3$–PMMA thermoelectric generators via dithiol-assisted conductivity and FEM-based geometry optimization

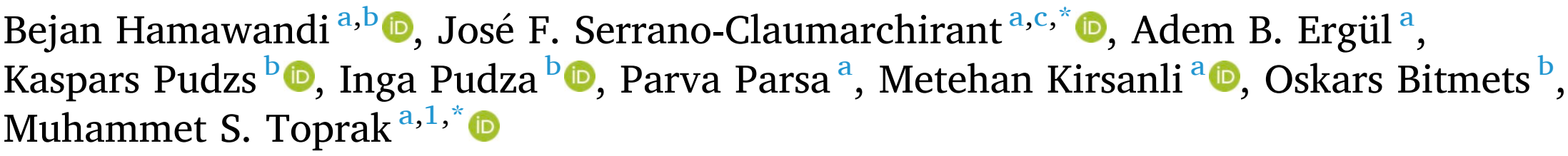

Bejan Hamawandi [a,b], José F. Serrano-Claumarchirant [a,c,*], Adem B. Ergül [a], Kaspars Pudzs [b], Inga Pudza [b], Parva Parsa [a], Metehan Kirsanli [a], Oskars Bitmets [b], Muhammet S. Toprak [a,1,*]

[a] Department of Applied Physics, KTH Royal Institute of Technology, SE-106 91 Stockholm, Sweden
[b] Institute of Solid State Physics, University of Latvia, LV-1063 Riga, Latvia
[c] Instituto de Tecnología Química, Universitat Politècnica València-Consejo Superior de Investigaciones Científicas, Av. dels Tarongers, València, 46022, Spain

A B S T R A C T

In recent decades, thermoelectric (TE) materials have proven to be a complementary source of renewable energy, as they can directly convert waste heat into electrical energy. Energy-efficient, reliable, and scalable synthetic routes for the fabrication of TE materials and their processing into functional devices via low-energy and low-waste routes are necessary for the broader adoption of these materials in various applications. In this work, we report the formulation of hybrid thermoelectric (*h*TE) inks based on nanostructured $Sb_2Te_3$ and $Bi_2Te_3$, using PMMA as the polymer matrix and hexanedithiol (HDT) as the binder. Percolation studies were conducted to determine the optimal film composition, with an 80% nanoparticle content yielding the highest TE performance. Finite element modelling (FEM) was employed to optimize the device geometry, including the cross-sectional area ratio of p- and n-type legs, to maximize power output. Based on these results, a flexible *h*TEG was fabricated using the optimized ink composition. The device exhibited an output power of 950 nW and a Power output Density (PoD) of 40.37 nW $cm^{-2}$ under a 30 K temperature gradient, significantly outperforming previously reported polymer-based flexible *h*TEGs incorporating chalcogenides. This study presents a sustainable and effective strategy for developing high-performance hybrid thermoelectric devices through ink formulation, composition optimization, and simulation-guided device design.

## 1. Introduction

Thermoelectric (TE) materials can directly convert temperature differences into electrical power, presenting a promising solution for waste heat recovery and reducing carbon footprints. The principle of thermoelectricity is based on thermopower, where the Seebeck effect generates an electric current when a temperature gradient is applied across a TE material. Their performance is defined by the dimensionless figure of merit ($zT$), which is the ratio between the electronic power factor ($S^2\sigma$) and thermal conductivity ($\kappa$) by the absolute temperature ($T$), expressed as $(S^2\sigma/\kappa)T$, where $S$ and $\sigma$ are the Seebeck coefficient and the electrical conductivity, respectively [1].

Bismuth and antimony telluride ($Bi_2Te_3$ and $Sb_2Te_3$) alloys exhibit superior TE properties up to about 150 °C, enabling effective waste heat recovery even at low-grade temperature gradients. Innovative fabrication techniques, including solution-chemical approaches and nanostructuring, have further improved the performance of TE materials and TE generators (TEGs) [2,3]. The deployment of TEGs in various applications, including automotive [4], manufacturing [5], power generation for the Internet of Things (IoT) demonstrates its potential to contribute to sustainable energy development.

In industrial processes, a substantial portion of energy is lost as waste heat, contributing to inefficiencies and a negative environmental impact [6]. TEGs offer a sustainable solution by harnessing waste heat and converting it into usable electrical energy, reducing the environmental impact of industrial activities. They operate without moving parts,

* Corresponding authors.
*E-mail addresses:* jserrano@kth.se (J.F. Serrano-Claumarchirant), toprak@kth.se (M.S. Toprak).
[1] Visiting Professor, Department of Physics, National University of Singapore, Singapore

making them reliable and maintenance-free. As research and development in TE technology continue to advance, the widespread adoption of TEGs could play a crucial role in achieving global carbon reduction goals [7]. Recent advancements in TE materials development and device engineering have considerably improved the performance and scalability of TEGs [5].

Hybrid TE (***h***TE) materials explore the possibility of combining inorganic TE materials with organic matrix-like polymers. A substantial body of literature exists on the advancements in organic-inorganic composites, highlighting the benefits of their combination, including cost-effectiveness, flexibility, scalability, and low thermal conductivity [8–11]. The combination of cost-effectiveness, scalability, and low thermal conductivity is achieved using organic components. Furthermore, high electrical conductivity, stability, and often material functionality, such as TE properties obtained from inorganic components, make this composite material perform efficiently in various low-grade temperatures and environmental conditions [12]. Due to their unique properties, ***h***TEs have shown significant potential for power generation and energy applications. Despite recent advancements in ***h***TEs, many challenges remain to be addressed for these composites to enhance both performance and practical applicability of ***h***TEs in next-generation TE generators (TEGs) [12,13].

Exploiting small organic molecules as constituents in electronic materials is currently being explored to develop semiconductor nanoparticle (NP)- based materials and devices[14,15]. Reports on the positive influence of organothiols (i.e., thiolated organic molecules) on the electronic properties of NP films exist. A nearly tenfold increase in mobility was recorded for semiconductor PbS films treated by ethane dithiol [16]. Recently, we demonstrated a similar trend of reduced resistance in organic-inorganic hybrid TE films using single thiol (DDT) and di-thiols (HDT) [17,18]. The interaction of organothiols with the NP surface was proposed to be redox-type, causing the reduction of surface oxide [19], which may also lead to reduced interfacial resistance between the particles.

Two significant factors that affect the effective design of TE devices in practical applications are power output and reliability. High power ensures effective energy conversion efficiency, broadening the devices' application scenarios. In addition to high stability, ensuring the reliability and long-term operation of TEGs in various environments is essential. Notably[20], computational simulations can effectively optimize the output power of devices by identifying the most significant parameters to optimize for maximum power output. Most of the work focuses mainly on simulating the optimal geometry of the TE legs. The influence of leg shape on TE performance under constant temperature and heat flux boundary conditions has been studied extensively, revealing that different leg geometries, cross-sections, and shapes can significantly impact the electrical and thermal performance of TE devices. Furthermore, investigations into thin-film TEGs have also highlighted the importance of leg shape, TEG configuration, and contact resistance in optimizing device performance [21,22]. Additionally, the impact of TE leg geometries on thermal resistance and power output has been demonstrated, providing compelling motivation to explore additive manufacturing of TE devices [23]. Earlier works on TE geometry and structure optimization have provided comprehensive insights into performance enhancement strategies, including the use of porous TE materials to reduce $\kappa$ without compromising electrical transport properties [24]. Novel numerical case studies have also been conducted to optimize the design of TE modules with hollow semiconductors, further enhancing their performance. Modeling and optimization analyses of hollow flexible-filler-based bulk TEGs for human body sensors have demonstrated the potential of these designs to improve output power and reduce fabrication costs [25,26].

In this work, we demonstrate *n*- and p-type TE materials synthesis and their processing into ***h***TEGs through bottom-up strategies. The positive effect of organothiols as binders on the transport performance of ***h***TE films is also evidenced. An ***h***TEG with multiple *n*- and p-type legs is fabricated using the as-synthesized $Bi_2Te_3$ and $Sb_2Te_3$ materials in an ink formulation, blended with PMMA as the polymer matrix. The power output of the ***h***TEG is optimized throughout the COMSOL simulation of the individual leg characteristics. The experimentally measured power output from the device is closely predicted by the simulation, demonstrating a case where theory can accurately predict the experimental performance of the TEG, suggesting the viability of its use for guiding the design of various ***h***TEG architectures for different application scenarios.

## 2. Experimental

### 2.1. Materials

All chemicals used in the synthesis and further processing were purchased from Sigma Aldrich (Stockholm, Sweden). They were used as received, with no further purification: Bismuth chloride ($BiCl_3$, 98% purity), Antimony chloride ($SbCl_3$, 99.95% purity), Tellurium powder (Te, 99.8% purity), Oleic acid ($C_{18}H_{34}O_2$), 1-Octadecene ($C_{18}H_{36}$, ODE), thioglycolic acid ($C_2H_4O_2S$, 98%, TGA), tri-butyl phosphine ($C_{12}H_{27}P$, 93.5%, TBP), 1,6-Hexanedithiol ($HS(CH_2)_6SH$, 96%, HDT), poly(methyl methacrylate) (PMMA, $M_w$~350,000 by GPC), 1-Methoxy-2-propyl acetate (MPA, 99.5%), methanol ($CH_3OH$, 99.8%) and n-hexane ($C_6H_{14}$, 95%). $BiCl_3$, $SbCl_3$, and Te powder were subsequently precursors for Bi, Sb, and Te ions. Oleic acid was used as a complexing agent for Bi and Sb precursors, ODE was used as the solvent, TBP as the complexing agent for Te, and TGA as the shape-directing agent. Methanol and hexane were used to wash the synthesized particles. MPA was used as a solvent for the PMMA solution and ink formulation.

### 2.2. Synthesis of $Bi_2Te_3$ and $Sb_2Te_3$

The synthesis was carried out using an MW-assisted thermolysis process, as detailed in our earlier work [27]. The details are given in the Supplementary Information.

### 2.3. Formulation of hybrid TE ink

The as-made powders of *n*- and p-type TE materials ($Bi_2Te_3$ and $Sb_2Te_3$, respectively) are used as fillers in the PMMA polymer matrix, with MPA (1-Methoxy-2-propyl acetate) as the solvent for the formulation of ***h***TE ink. Polyethylene terephthalate (PET) substrates were precleaned utilizing a sequence of ultrasonication with acetone, isopropanol, and deionized water. An appropriate amount of PMMA was dissolved in MPA by heating the mixture for 12 min at 100 °C using a microwave (MW) power of 400 W under constant stirring via a MW synthesizer (Biotage Initiator+) to obtain a 12 wt% PMMA+MPA clear solution. In the next step, an appropriate amount of $Bi_2Te_3$ or $Sb_2Te_3$ was added to 0.5 mL of the solutions above, and the mixture was stirred for 30 min at 80 °C. This was followed by 5 min of ultrasonication and an additional 3 min of mixing on a hot plate to produce homogeneous inks containing 60 to 80 wt% NPs. Two sets of samples were fabricated: one with and one without the addition of a binder molecule, HDT, 5 wt% relative to the weight of the NP filler ($Bi_2Te_3$/$Sb_2Te_3$), to investigate its effect on the electronic performance of the ***h***TE films. The ***h***TE were fabricated by casting 200 μL of the formulated $Bi_2Te_3(Sb_2Te_3)$/PMMA ink within the confined area created by Kapton tape on a PET substrate (1.60 $cm^2$), subsequently dried at 80 °C for 1 hour under ambient conditions. The final ***h***TEG device, designed with an optimized geometry through Finite Element Modeling (FEM), was manufactured following the same methodology previously described. To maintain a constant leg thickness, the volume of ink deposited per unit area was kept constant. For performance comparison, another ***h***TEG device with the same number of legs was fabricated using a symmetric leg configuration. Finally, conductive silver paint was applied to the dried hybrid film utilizing a brush to connect the individual legs into p-n junctions. Fig. 1 illustrates the schematics of the processing of TE ink into ***h***TE films. The rationale behind the selection of PMMA as the matrix, MPA as the solvent, and HDT as the binder is based on the evaluation of the electrical conductivity of $Sb_2Te_3$ and $Bi_2Te_3$ films prepared with different polymer matrices, solvents, and binders (see **Supplementary Fig. 1**).

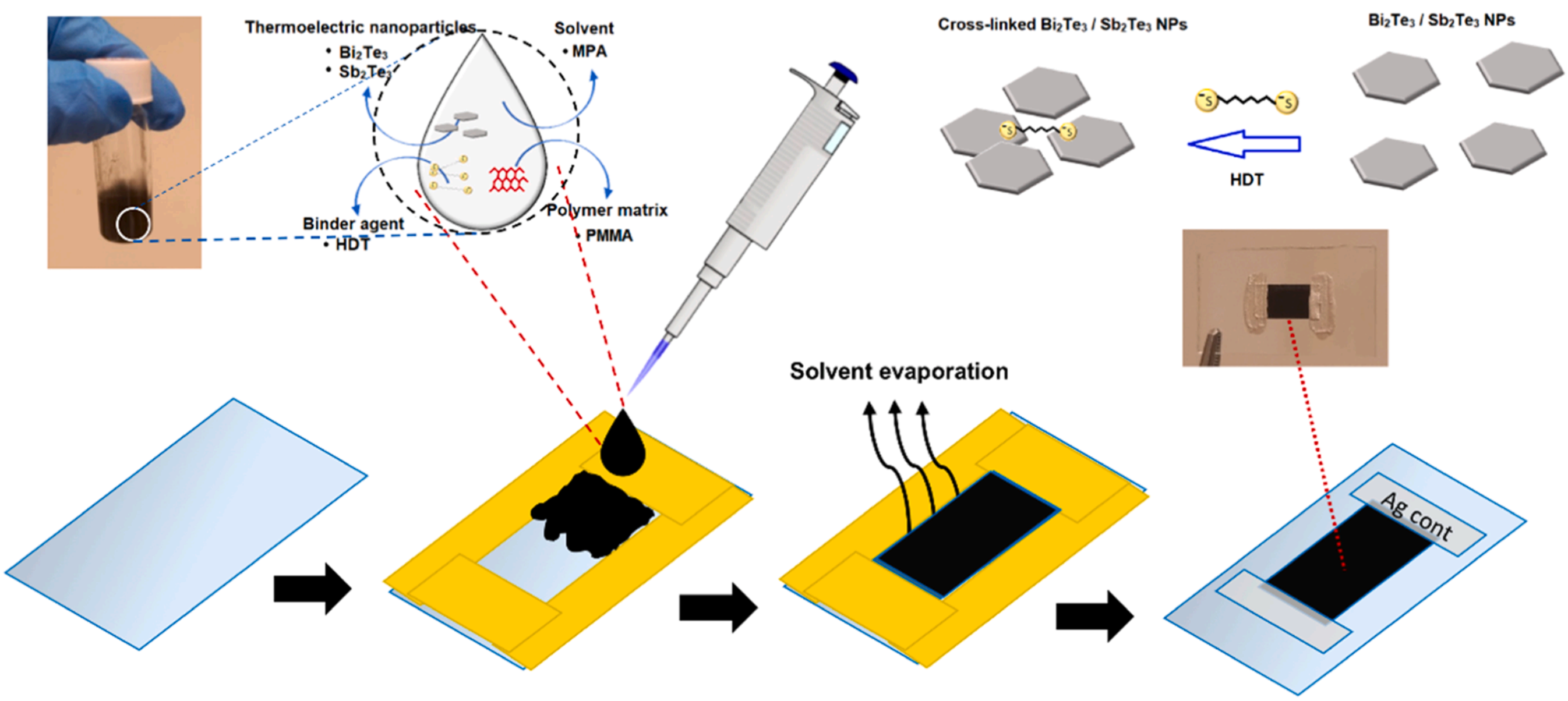

**Fig. 1.** Schematics of processing of TE ink for the fabrication of ***h***TE films.

### *2.4. Simulations*

The geometry optimization of TEG was modeled using FEM, employing COMSOL Multiphysics software (v6.0). The simulations were carried out under steady-state conditions usingThe program utilizes Heat Transfer and AC/DC Modules (electrical current and electrical circuits) for a steady-state (stationary) analysis. Ground and Terminal connections were integrated into the TEG, and a variable resistor was attached to complete the circuit, representing the external load resistance. For the thermal model, two Dirichlet boundary conditions were applied to the model on the Heat Transfer module. The temperature of the front side of the PET substrate was maintained at 300 K, and the temperature of the back side of the substrate was maintained at 310 K during the geometry optimization. These boundary conditions define a constant temperature gradient across the device, enabling a direct comparison of the effect of geometry on the thermoelectric performance during the optimization process. No external heat flux was imposed, as the thermal driving force is fully determined by the prescribed temperature difference. Under these assumptions, the temperature distribution and resulting electrical output are governed by the coupled heat transfer and electrical transport equations. It should be noted that the use of fixed temperatures represents an idealized condition. In practical operation, the cold side is typically influenced by convective heat exchange with the surrounding environment, which may reduce the effective temperature gradient. Therefore, the simulated results correspond to an upper-bound estimation of device performance under controlled thermal conditions.

The parametric Solver was configured for the Automatic (Newton) method using a physics-controlled mesh with a finer element size. The physical constants of the materials employed were taken from the COMSOL Materials database. However, the transport properties ($\sigma$ and $S$) of the ***h***TE films of $Bi_2Te_3$ ($Sb_2Te_3$)/PMMA corresponding to the individual legs were obtained from the experimental measurements.

### *2.5. Characterization of TE materials, hTE films, and hTEGs*

Crystalline phases and phase purity of as-synthesized TE materials were investigated using X-ray powder diffraction (XRPD) with a Philips PANalytical X'Pert Pro Powder Diffractometer, employing Cu-Kα1 radiation and a scan speed of 0.04°/s in continuous scan mode.

Synchrotron radiation X-ray absorption spectroscopy (XAS) measurements were performed at the P65 Applied XAS beamline of PETRA III, DESY[28](see **Supplementary Material**). Morphology and microstructure of the powder TE materials and the ***h***TE films were performed using Electron Microscopy (SEM (FEI Nova 200) and TEM (JEM-2100F, 200 kV, JEOL). A 200-mesh copper grid was used for TEM analysis, where 100 μL of the sample was pipetted onto it and then air-dried before analysis. The thickness of the films was measured using a profilometer (Profiler KLA-Tencor P15, CA, USA), while the lateral dimension was pre-determined using the Kapton tape template.

Electronic transport properties of the ***h***TE films were evaluated by measuring their electrical conductivity ($\sigma$) and the Seebeck coefficient ($S$) of three independent samples for each composition. The $\sigma$ was determined using the linear four-point probe method. Four equidistant, rounded probes were applied in the center of the film surface, and a current was applied between the outer probes, measuring the voltage between the inner ones. Using the IV curve and the film thickness ($d$), the electrical resistivity ($\rho$) was calculated and converted into $\sigma$ (see **Supplementary Material**). Hall mobility measurements were conducted using a PPMS DynaCool system to determine both carrier concentration and mobility. A $3 \times 5$ mm film section was cut and contacted using silver paste and copper wires, following the configuration shown in **Supplementary Fig. 2** for the dual sample holder. A constant current of 0.5 mA was applied, and the voltage was recorded across a range of magnetic field values, from −1000 to 1000 Oe in 500 steps. From the resulting data, the Hall coefficient ($R_H$), carrier concentration ($n$), and carrier mobility ($\mu$) were determined.

## 3. Results and discussion

### *3.1. Characterization of as-synthesized TE material*

$Sb_2Te_3$ and $Bi_2Te_3$ materials, as p- and n-type TE materials, have been synthesized in large quantities via an energy- and time-efficient microwave-assisted thermolysis route, with a reaction yield of 98%.

Structural analysis was performed using XRPD on the as-synthesized powders, and the results are presented in **Supplementary Fig. 3**, along with the corresponding Miller indices. The diffraction patterns are normalized to the most intense diffraction peak, indexed to the (015) plane. The crystalline phases were identified as $Sb_2Te_3$ (ICDD: 96–900–7591) and $Bi_2Te_3$ (ICDD: 01–085–0439), exhibiting a rhombohedral crystal structure. The sharp and high-intensity diffraction peaks are a good indication of the high crystallinity of the synthesized materials.

The experimental extended X-ray absorption fine structure (EXAFS) spectra ($\chi(k)k^2$) and corresponding Fourier transforms for the Sb and Te K-edges in $Sb_2Te_3$, and the Bi $L_3$-edge and Te K-edge in $Bi_2Te_3$ powder samples, recorded at 10 and 300 K (**Supplementary Fig. 4**), show a pronounced reduction in amplitude at higher temperatures, indicating increased thermal disorder due to weak interatomic bonding. The data were analyzed using reverse Monte Carlo (RMC) simulations[27], resulting in structural configurations consistent with the rhombohedral crystal structure. The extracted mean-square relative displacement (MSRD) factors and effective interatomic force constants highlight the strong anisotropy of thermal conductivity inherent to the layered architectures of $Bi_2Te_3$ and $Sb_2Te_3$.

TEM analysis was performed on the as-made materials, and a few micrographs of the samples are presented in Fig. 2. Samples lying along the a-b axes are very thin and are transparent in the TEM analysis. A nearly hexagonal platelet morphology can also be observed in the TEM micrographs.

### 3.2. Evaluations of hybrid TE films

#### 3.2.1. Microstructure analysis of the films

Microstructural and morphological analyses of the as-made powders have been reported earlier using SEM [27]. Both samples exhibit a hexagonal platelet morphology, displaying significantly different lateral dimensions (within the range of 20–500 nm for $Bi_2Te_3$ and 20–4000 nm for $Sb_2Te_3$) [27]. SEM micrographs of typical hybrid films in the PMMA matrix at different magnifications are presented in Fig. 3. Fig. 3**a,b** shows $Bi_2Te_3$ – PMMA, and Fig. 3**c,d** shows $Sb_2Te_3$ ***h***TE films, where the morphology of constituent nanoparticles can be clearly seen.

The cross-sectional view of the film containing $Sb_2Te_3$ NPs is shown in Fig. 3**d** The micrograph shows a film thickness of around 50 μm with the NPs homogeneously blended and distributed within the PMMA polymer matrix. The passive top layer polymer protects the filler particles against air exposure and environmental conditions. It is one of the primary concerns for the long-term stability of ***h***TEs, as these factors can lead to the oxidation of TE materials and, consequently, a gradual decline in their performance. PMMA is a hydrophobic polymer that is air-impermeable and thermally stable up to 150 °C, providing a thermally stable and protective matrix for the embedded NPs [29–32].

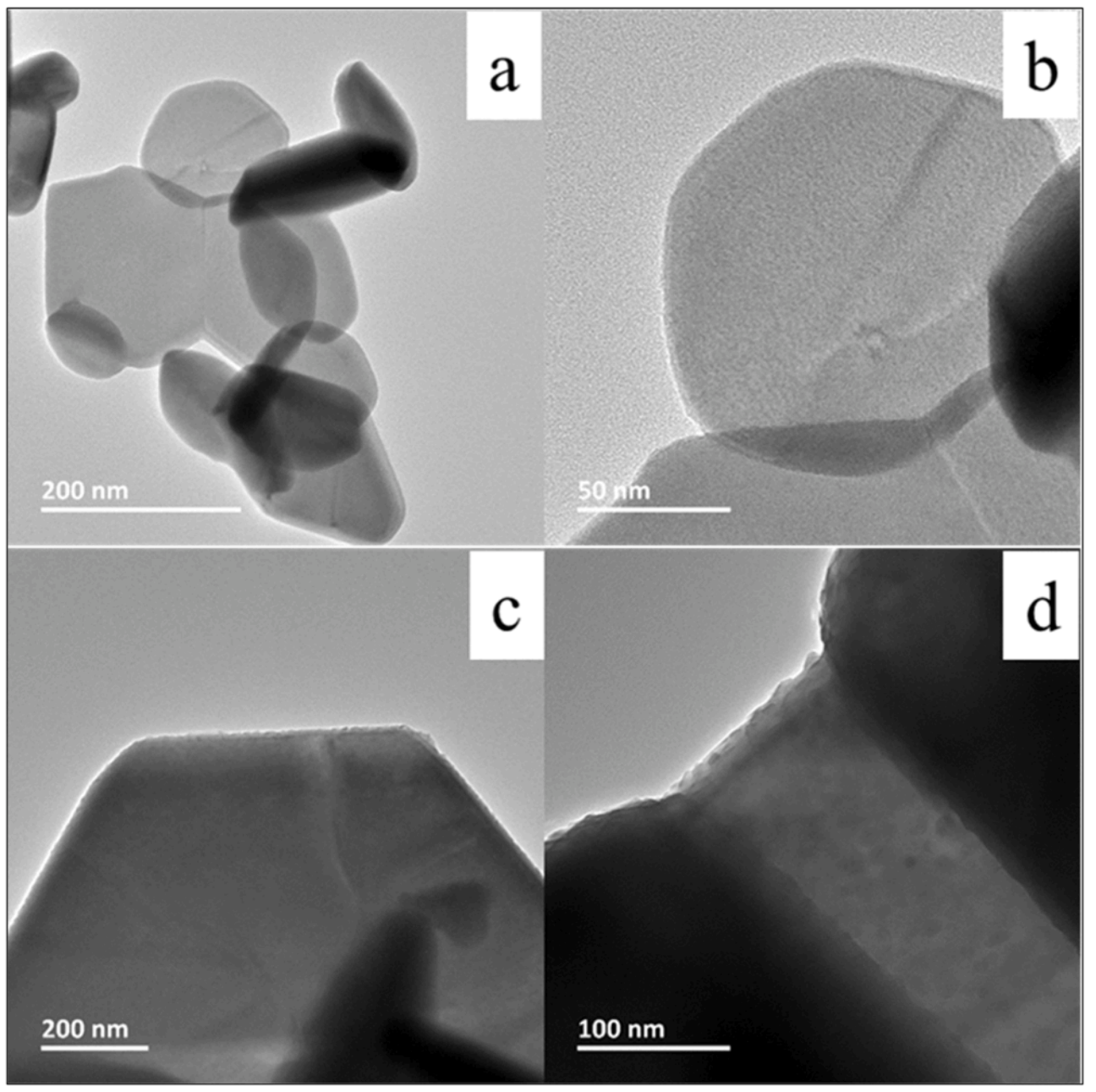

**Fig. 2.** TEM micrographs of as-synthesized TE materials (a, b) $Bi_2Te_3$, and (c, d) $Sb_2Te_3$ through MW-assisted thermolysis.

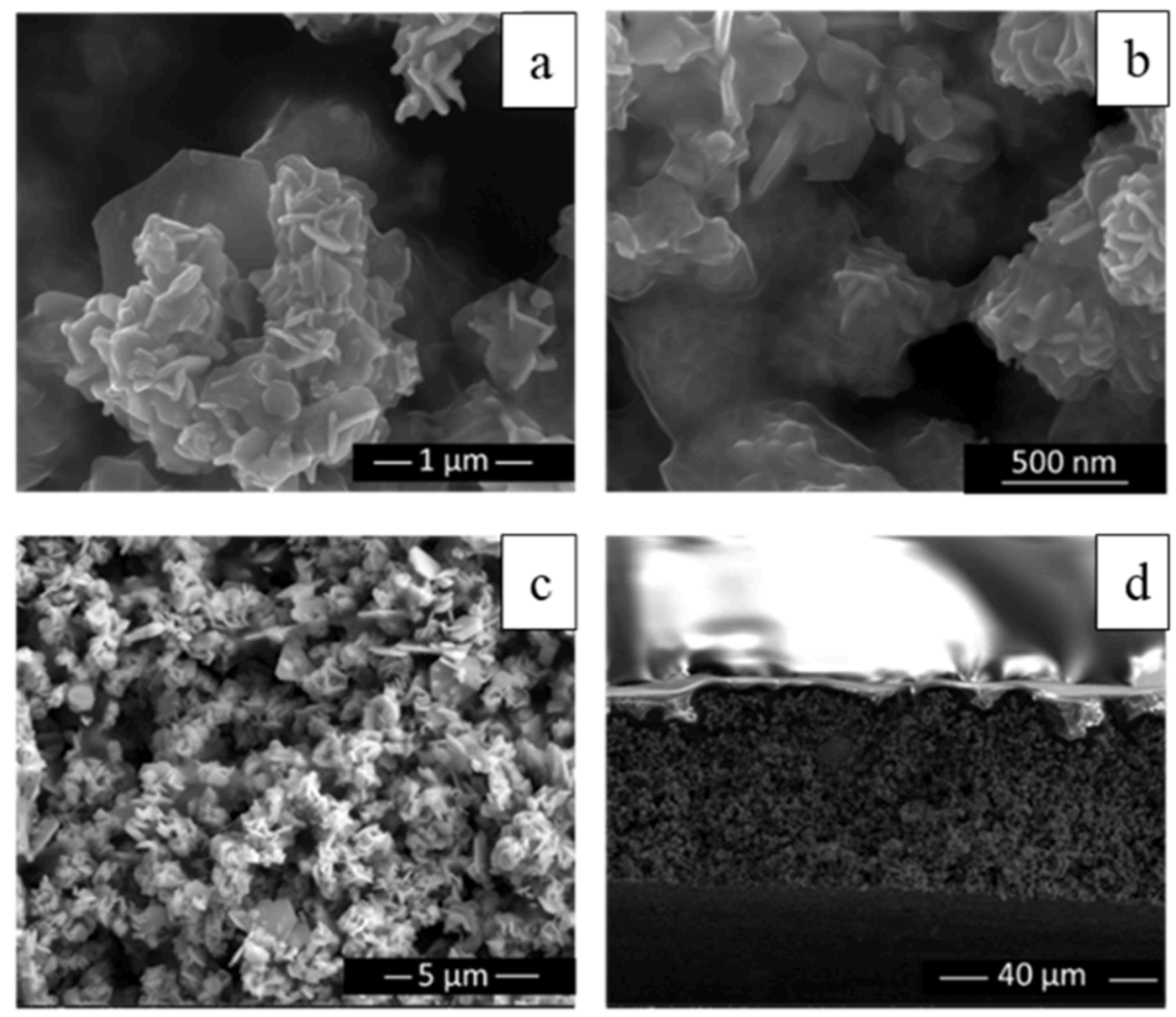

**Fig. 3.** SEM micrographs of ***h***TE films at different magnifications: (a-b) $Bi_2Te_3$-PMMA, (c-d) $Sb_2Te_3$-PMMA, where (d) shows a typical cross-sectional view of $Sb_2Te_3$–PMMA hybrid film.

### 3.3. Transport properties of hTE films

Seebeck coefficient (***S***) and resistance measurements were performed on ***h***TE films with varying nanoparticle content (60, 70, and 80%, corresponding to PMMA polymer contents of 40, 30, and 20%, respectively) under two conditions: binder-free and with the addition of 5 wt% HDT as the binder. It has been previously demonstrated that organothiols can positively influence electronic transport in films composed of nanoparticulate materials [17]. The effect of adding a binder molecule to the ink was investigated to study its influence on the carrier mobility and conductivity of the ***h***TE films. The complete percolation curve for the $Sb_2Te_3$/PMMA and $Bi_2Te_3$/PMMA ink formulation with and without HDT can be found in **Supplementary Fig. 5**.

The samples' $\boldsymbol{\sigma}$ was calculated based on the equation for linear four-point probe electrical resistance ($\boldsymbol{\sigma} = \ln 2/\pi\, \boldsymbol{R}\, \boldsymbol{t}$), where ***R*** represents the resistance measured with linear 4- probe, and ***t*** is the film thickness. The resulting values of $\boldsymbol{\sigma}$ for the samples are presented in Fig. 4 and are summarized in **Supplementary Table 1** for all samples. The dimensions of the films applied to the PET substrate were as follows: a length of 1.5 cm, a width of 0.7 cm, and a thickness of 50 μm. The ***h***TE films show an increasing trend of $\boldsymbol{\sigma}$ with increasing NP content (Fig. 4**(a)**). $Sb_2Te_3$ ***h***TE films without HDT show a $\boldsymbol{\sigma}$ value of 0.57 S/cm (60%), increasing to 4.2 S/cm (70%) and finally to 5.3 S/cm (80%) with increasing NP content. Upon the addition of HDT to the formulation of the inks, the $\boldsymbol{\sigma}$ values increase significantly from 3.1 S/cm (60%) to 6.0 S/cm (70%) and finally to 8.1 S/cm (80%). The same trends are observed for $Bi_2Te_3$ ***h***TE films upon the addition of HDT, where the $\boldsymbol{\sigma}$ values increase from 0.1 to 0.3 S/cm, 0.7 to 2.1 S/cm, and 0.8 to 3.2 S/cm for the films with 60%, 70%, and 80% NP content, respectively. There is an apparent positive impact of HDT addition on the $\boldsymbol{\sigma}$, which will be discussed later, along with the carrier concentration and Hall mobility measurements. ***S*** values of the ***h***TE films were also evaluated for films with the same NP content; the results are presented in Fig. 4**b** and summarized in **Supplementary Table 1** for all samples. $Sb_2Te_3$ ***h***TE films exhibit positive ***S*** values – indicative of p-type transport - in the range of approximately 200–225 μV/K, with an incremental increase upon the addition of HDT. $Bi_2Te_3$ films exhibit negative ***S*** values, indicative of n-type transport, ranging from −150 to −175 μV/K, which increase slightly in absolute value to −175 to −210 μV/K upon the addition of HDT. The slight increase in ***S*** values upon HDT addition can be attributed to an increase in carrier mobility. Finally, using the values of *σ* and ***S***, the electronic power factor ($S^2\sigma$) was estimated and presented in Fig. 4**(c)**. In agreement with the trends of *σ* and ***S***, the power factor (**PF**) values increase with increasing NP content, with a further improvement in the presence of HDT for both *n*- and p-type ***h***TE films. The highest **PF** values were estimated for ***h***TE films with the highest NP content of 80%. The highest **PF** value of 40 μW $m^{-1}$ $K^{-2}$ has been calculated for $Sb_2Te_3$, while the highest value of about 14.1 μW $m^{-1}$ $K^{-2}$ has been obtained for $Bi_2Te_3$ ***h***TE films.

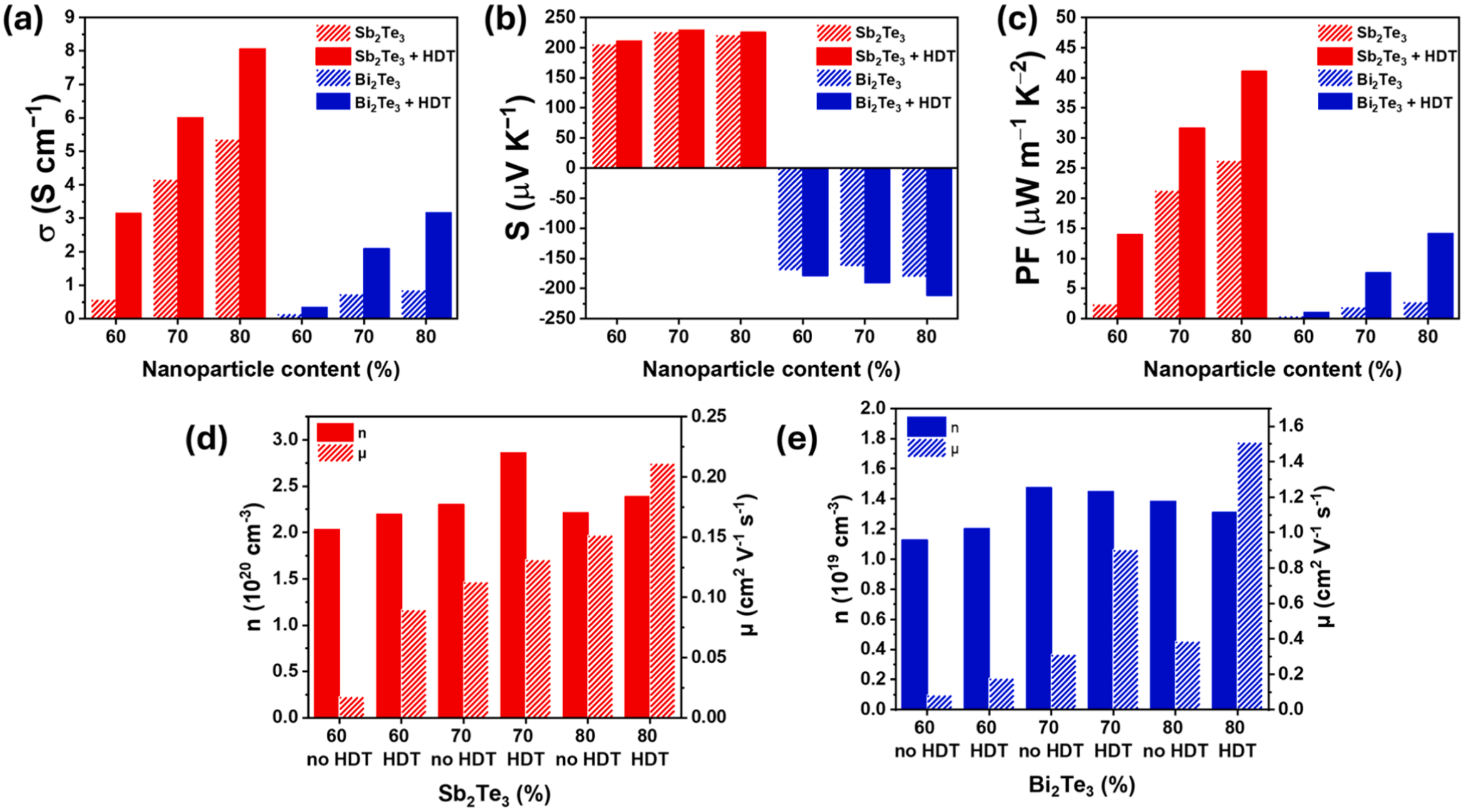

**Fig. 4.** Electronic transport properties of the as-made ***h***TE films with different filler ratios (60–80%) of $Bi_2Te_3$ and $Sb_2Te_3$ NP in the polymer matrix, in the absence and presence of with and without HDT; (a) electrical conductivity (*σ*), (b) Seebeck coefficient (***S***), and (c) electrical power factor (**PF**); carrier concentration (**n**) and Hall mobility (**μ**) of (d) $Sb_2Te_3$ and (e) $Bi_2Te_3$.

The samples with organothiol HDT in their compositions (Fig. 4**c**) exhibit elevated **PF** compared to their HDT-free counterparts. This increase is attributed to a significant increase in their conductivity (*σ*). To shed light on the real effect of the organothiol addition, we performed Hall mobility measurements on the characterized ***h***TE films. Fig. 4**d** and 4**e** show the obtained values for carrier concentration (***n***) and Hall mobility (***μ***) for $Sb_2Te_3$ and $Bi_2Te_3$ ***h***TE films with different NP content, in the absence and presence of HDT. A general trend of slightly increased ***n*** with increasing NP content is observed for both samples across all compositions. For a given composition of ***h***TE films, ***n*** is of the same order of magnitude, and it does not exhibit a strong dependency on the HDT addition, whereas the ***μ*** values show a significant jump in the presence of HDT. This reveals that HDT reduces the interface resistance between the NPs and the polymer matrix in ***h***TE films, improving the ***μ*** and *σ* values. This is consistent with percolation curve studies (see **Supplementary Fig. 5**), which show that adding HDT reduces the percolation threshold and improves the prefactor governing the efficiency of charge transport above the threshold. This specifically enhances the **PF** of the ***h***TE films. These results align with previous works, where organothiol was used to reduce the interface resistance between particles [17,18,33]. Bifunctional thiols can replace long insulating surface ligands, such as those from the oleic acid solvent, and form direct molecular bridges between neighboring particles. This reduces interparticle spacing and it may lower the tunneling barrier for charge carriers. This mechanism is well established in nanoparticle films, where short-chain ligands enhance electronic coupling and charge transport. Other work on alkane dithiol-linked junctions shows that HDT shortens effective tunneling distances and increases current at nanoparticle–thiol interfaces, an effect attributed to stronger metal–sulfur bonds and improved electron coupling. Consequently, the observed increase in mobility is consistent with a reduction in interfacial resistance resulting from improved electronic coupling between particles [34–36].

Our findings show the following experimental trends: *σ* increases strongly, mobility increases, and the absolute value of ***S*** increases. While this combination is consistent with more efficient charge transport pathways induced by HDT treatment, it does not uniquely identify a single dominant mechanism. In addition to enhanced carrier mobility, several concurrent effects may contribute to the observed behavior, including modifications of interparticle tunneling barriers, variations in effective carrier concentration, changes in surface states or trap density, or alterations in the energy-dependent transport near the Fermi level. In this context, interfacial phenomena, such as energy-dependent transmission across nanoparticle junctions, could lead to signatures resembling energy filtering, with a preferential contribution of higher-energy carriers and a corresponding slight increase in ***S*** [37,38].

This combination improves the power factor $S^2\sigma$, which is a result of synergy we sought to achieve through the ***h***TE strategy. Based on the electronic transport property data, ***h***TE formulations with 80% NP ($Sb_2Te_3$ or $Bi_2Te_3$) content were chosen and used for the simulation and fabrication of ***h***TEG in the following sections.

### 3.4. Simulation, optimization and fabrication of hTEG devices

Once the single ***h***TE legs were characterized individually, their ***σ*** and ***S*** values were input into the model to determine the optimal geometry. The modeling of ***h***TEG utilized the properties of ***h***TE films containing 80% NP content. As shown in Fig. 4**a**, the ***σ*** of $Sb_2Te_3$ ***h***TE films (p-leg) is twice that of the $Bi_2Te_3$ ***h***TE films (n-leg). Consequently, the first step in optimizing the geometry of the ***h***TEG involves identifying the optimal cross-sectional area ratio between the p-leg and the n-leg (p:n). For this purpose, the width of the p-leg was fixed at 7 mm, while the width of the n-leg was varied, maintaining a constant leg length and applying a temperature gradient of 10 K.

Since increasing the cross-sectional area reduces the device's internal resistance, the power output increases with a larger cross-sectional area, potentially resulting in an excessively large TEG. Therefore, geometry optimization was conducted based on the power output density (***PoD***), enabling the identification of the geometry that delivers the highest power output per unit area while minimizing the device's footprint. The results of this initial simulation step are shown in Fig. 5**a**. As expected, increasing the width of the n-leg leads to a reduction in the device's internal resistance. Furthermore, it was observed that a p:n ratio of 1:2 yields the highest ***PoD***. Thus, the width of the n-leg should always be double that of the p-leg, consistent with the measured ***σ*** values.

In the next step, the optimal width of the p-leg was evaluated. A parametric study was conducted in which the width of the p-leg was varied while maintaining a constant p:n ratio of 1:2, a leg length of 15 mm, and a temperature difference of 10 K. The results, shown in Fig. 5**b**, indicate that the power output density increases gradually with the p-leg width until it reaches 14 mm, where a saturation point is observed (the differences between 13 and 14 mm are negligible). As in the first step, increasing the cross-sectional area reduces the TEG's internal resistance.

Finally, the optimization of the leg length was addressed. The p-leg width was fixed at 14 mm, the n-leg width at 28 mm, and the leg length was varied between 10 and 30 mm, with a thermal gradient of 10 K applied. Leg lengths shorter than 10 mm were excluded from the study due to the increased difficulty of achieving high thermal gradients. In this case, the maximum ***PoD*** was obtained with short legs (10 mm, Fig. 5**c**). Additionally, the internal resistance decreased slightly from

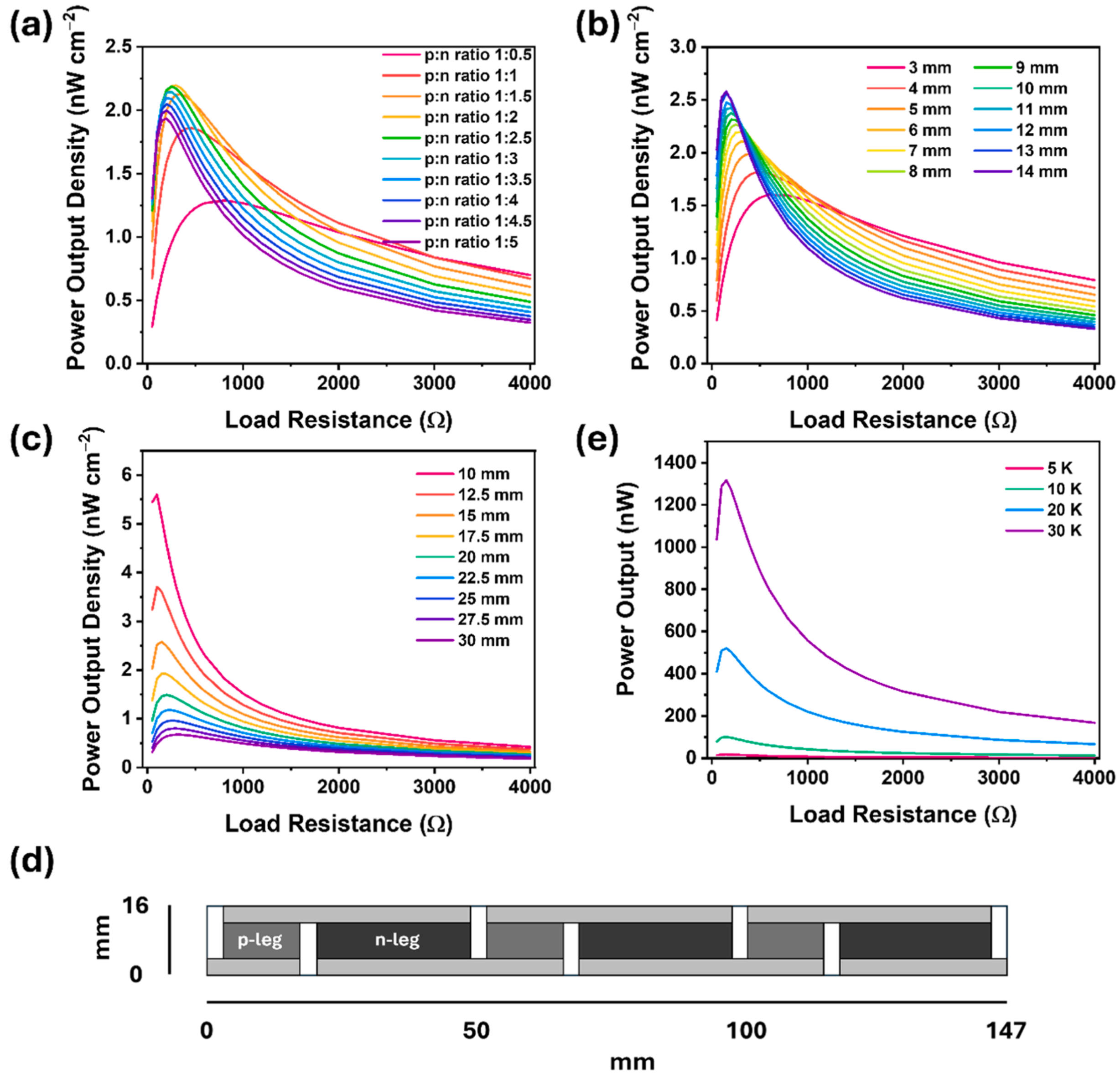

**Fig. 5.** Finite element modeling (FEM) optimization of (a) p:n width ratio, (b) p-leg width, and (c) leg length as a function of power output density -***PoD***. (d) Schematic of the TEG final geometry - The three legs are marked as "n" and the three legs as "p", representing the NPs content of the legs as $Bi_2Te_3$ and $Sb_2Te_3$, respectively. (e) Power Output as a Function of thermal gradient, $\Delta T$.

250 ohms (length of 30 mm) to 150 ohms (length of 10 mm).

Thus, the optimal geometry for a TEG using the formulated inks in this study is established as rectangles with a length of 10 mm and a width of 14 mm for the p-leg and 28 mm for the n-leg, as illustrated in Fig. 5**d** Finally, the optimized geometry simulated the TEG's power output under different thermal gradients (Fig. 5**e**). The simulation predicts that under a $\Delta T$ of 30 K, a power output of 1.3 μW could be achieved.

The demonstrator ***h***TEG device was fabricated employing the geometry optimized by FEM. During the optimization process, the thickness of the legs was maintained by keeping the volume of hybrid ink deposited per unit area constant at 125 μL cm$^{-2}$. The same volumetric factor was used for the previously characterized individual legs. Thus, each p-type and n-type leg was obtained with 175 μL and 350 μL of the formulated hybrid-ink, which implies a total of 0.756 g of NPs (0.252 g $Sb_2Te_3$ and 0.504 g $Bi_2Te_3$ in total for the p and n-legs, respectively) for the complete ***h***TEG fabrication. Fig. 6**a** illustrates the ***h***TEG with three p-type and three n-type ***h***TE legs linked at their ends using silver cement, forming three thermocouples. An important parameter to evaluate during the development of TEG is the variation of internal resistance as a function of the number of legs. Given that the cross-sectional area of both p-type and n-type legs has been optimized, a linear evolution of resistance with the number of legs is expected, as illustrated in Fig. 6**b** The internal resistance of the ***h***TEG exhibits an almost linear increase from 18.2 Ω to 186.7 Ω. However, the final internal resistance of the optimized ***h***TEG is slightly higher than that of the simulated ***h***TEG (151.9 Ω), likely due to contact resistance at the interface between the legs and the silver cement electrode. In order to verify this, Transfer Length Method (TLM) measurements were performed to determine the contact resistance between the legs and the silver paste (see **Supplementary Fig. 6**). For this purpose, the distance between the silver paste electrodes varied from 2 to 10 mm, and by extrapolating to 0 mm, the contact resistances of silver paste/$Sb_2Te_3$ and silver paste/$Bi_2Te_3$ were found to be 2.7 Ω and 2.8 Ω, respectively. Considering the number of silver paste interfaces with the p- and n-type legs and adding these values to the theoretical internal resistance of the device, the resulting internal resistance is close to the experimental value (185.0 Ω). This indicates that the main reason for the increased internal resistance compared to theoretical value arises from the contact resistance at the silver paste/leg interfaces.

Subsequently, the ***h***TEG was subjected to various constant thermal gradients ($\Delta T = 5, 10, 20, 30$ K), during which different load resistances ranging from 10 Ω to 4000 Ω were applied. The voltage was measured, and using the voltage and load resistance values, the power output was calculated. The results are presented in Fig. 6**c**, which shows that both the power output and voltage increase with temperature.

**Supplementary Figs. 6** and **7** provides a magnified view of the power output values (not visible under low thermal gradients) and a fit of the output voltage using the following equation:

$$V = V_{OC}\frac{R_L}{R_I + R_L}$$

where $V_{OC}$ is the open-circuit voltage, $R_L$ is the load resistance, and $R_I$ is the internal resistance of the device. From this fitting, the internal resistance of the ***h***TEG and the $V_{OC}$ can be determined. Table 1 summarizes the experimental results obtained for the optimized TEG, including the derived parameters. The optimized ***h***TEG, fabricated using <1 g of TE powders, shows around 1 μW power output. As synthetic and processing technologies are easily scalable, this may enable a broader entry of TEs applied to larger areas.

Table 2 shows the comparative analysis of flexible ***h***TEGs, which reveals three main categories of materials: (i) carbon nanotube–based composites with conducting polymers or chalcogenides, (ii) chalcogenides combined with conducting polymers, and (iii) chalcogenides embedded in insulating polymer matrices. Each category displays distinct advantages and limitations in terms of electrical transport, mechanical flexibility, and power output. Carbon nanotube (CNT)–based composites, such as SWCNTs/PVP and PEDOT:PSS/SWCNT, achieve the highest reported power outputs and power output densities, reaching values in the μW cm$^{-2}$. This superior performance originates from the intrinsically high electrical conductivity and percolative transport pathways provided by CNTs [39]. Nevertheless, their practical application is often constrained by high processing costs, difficulties in large-area integration, and limited material availability [40,41].

**Table 1**
Internal resistance ($R_I$), maximum power output ($P_{max}$), maximum power output density ($PoD_{max}$), and open-circuit voltage ($V_{OC}$) for the TEG measured at different thermal gradients, $\Delta T$.

| $\Delta T$ | $R_I$ (Ω) | $P_{max}$ (nW) | $PoD_{max}$ (nW cm$^{-2}$) | $V_{OC}$ (mV) |
|---|---|---|---|---|
| 5 K | 184.13 | 13.65 | 0.58 | 3.00 |
| 10 K | 180.49 | 52.22 | 2.22 | 6.00 |
| 20 K | 180.66 | 361.27 | 15.36 | 16.02 |
| 30 K | 179.61 | 949.52 | 40.37 | 25.99 |

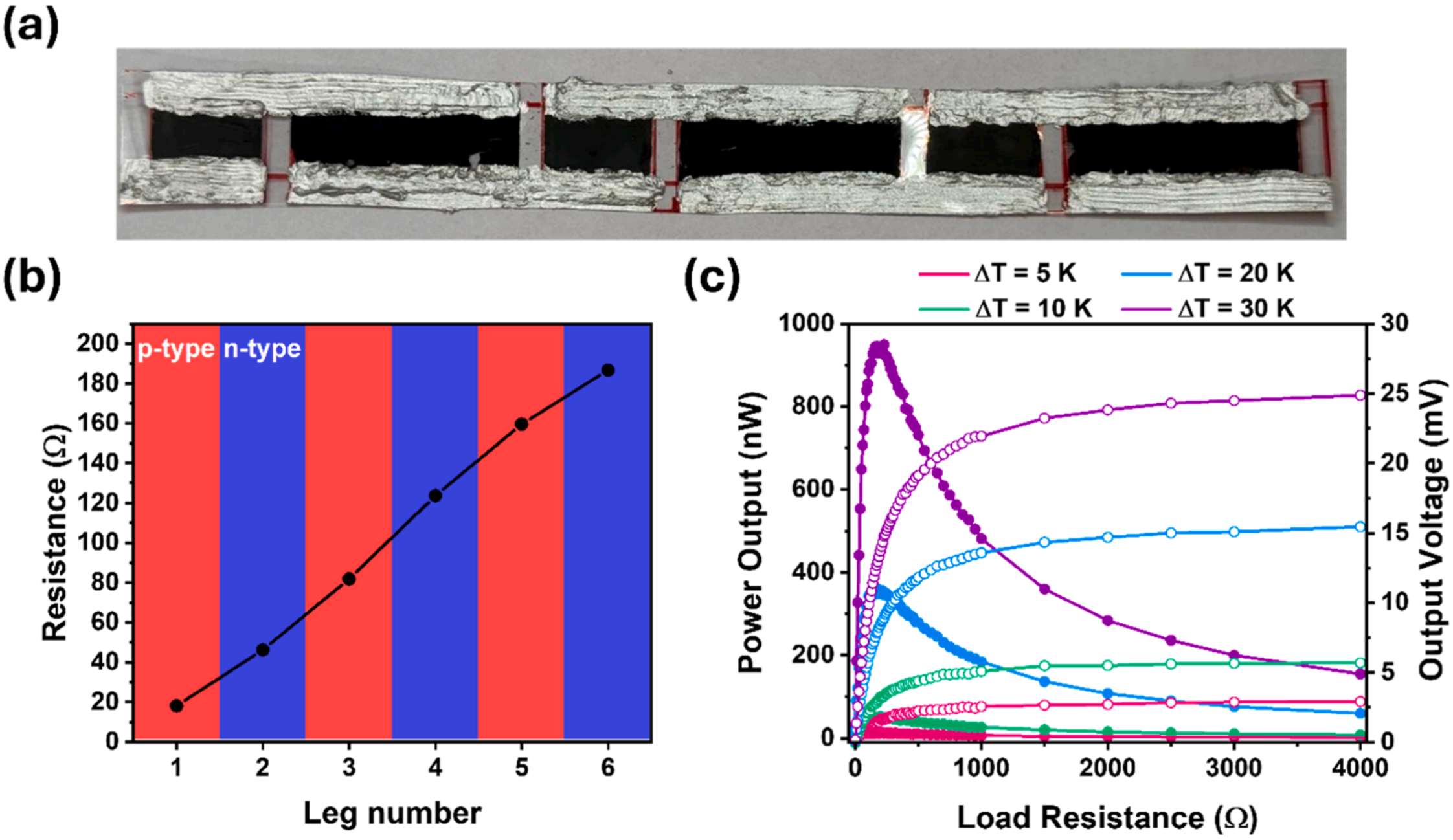

**Fig. 6.** (a) Photograph of the optimized ***h***TEG device, with 6 ***h***TE (80% NPs + 20% PMMA) legs (3 p-n junctions) on a PET substrate connected via silver cement at their ends. (b) Measured internal resistance change per leg throughout the device. (c) The ***h***TEG's power output and output voltage are under various load resistances and temperature gradients (● closed dots for Power Output, ◦ open dots for Output Voltage).

**Table 2**
Comparative summary of flexible ***h***TEGs reported in the literature, highlighting the hybrid TE material and architecture, number of thermoelectric legs, maximum output power ($P_{max}$), power output density (***PoD***), and applied temperature difference ($\Delta T$).

| Hybrid TE material and architecture | N° legs | $P_{max}$ (nW) | *PoD* (nW cm$^{-2}$) | $\Delta T$ (°C) | Reference |
|---|---|---|---|---|---|
| MWCNT-$NH_2$/NiO (p-p architecture) | 15 | 1.44 | 59.3 | 100 | [42] |
| SWCNTs/PVP (p-n architecture) | 6 | 1440 | 400,000 | 56 | [39] |
| PEDOT:PSS/SWCNT SWCNT:PEI (p-n architecture) | 18 | 2600 | 361,000 | 48 | [43] |
| $Ag_2Se$/SWCNT (n-n architecture) | 6 | 725 | - | 30 | [44] |
| $Bi_{0.5}Sb_{1.5}Te_3$/PEDOT:PSS (p-p architecture) | 6 | 45 | - | 25 | [45] |
| $Cu_3Se_2$/PANI-$H_2SO_4$ (p-p architecture) | 8 | 25.69 | 2380 | 100 | [46] |
| $Bi_2Te_{2.7}Se_{0.3}$/Silk fibroin $Bi_{0.5}Sb_{1.5}Te_3$/Silk fibroin (p-n architecture) | 6 | 20 | - | 40 | [29] |
| $Bi_2Te_3$/PVDF (p-n architecture) | 6 | 12 | - | 25 | [30] |
| $Sb_2Te_3$/HDT/PMMA $Bi_2Te_3$/HDT/PMMA (p-n architecture) | 6 | 949.52 | 40.37 | 30 | This work |

In contrast, ***h***TEGs constructed from chalcogenides and conducting polymers, such as $Cu_3Se_2$/PANI:$H_2SO_4$ or $Bi_{0.5}Sb_{1.5}Te_3$/PEDOT:PSS, exhibit considerably lower performance, typically within tens to hundreds of nW of power output. This reduction is attributed to the limited electrical conductivity of the polymer matrices, as well as interfacial mismatches between the organic and inorganic components, which hinder effective charge transport [40,41]. A similar trend is observed for chalcogenides dispersed in insulating polymer matrices, such as $Bi_2Te_3$/PVDF or $Bi_2Te_{2.7}Se_{0.3}$/silk fibroin, where the insulating nature of the host polymer severely limits electron transport, resulting in maximum power outputs below 25 nW. Within this context, the results of the present work ($Sb_2Te_3$/HDT/PMMA and $Bi_2Te_3$/HDT/PMMA) demonstrate a significant advancement for the category of chalcogenide–insulating polymer composites. The device achieves a maximum output power of 949.52 nW at $\Delta T = 30$ °C, representing an improvement of more than one order of magnitude compared to previously reported systems employing insulating matrices (12–25 nW). Although the corresponding power output density (40.37 nW cm$^{-2}$) is still far below the values obtained with CNT-based composites, it remains competitive when compared with other polymer-based systems and is particularly remarkable considering the relatively modest temperature gradient applied. This suggests that enhanced performance stems from the optimized hybrid architecture, rather than from extreme thermal conditions.

One of the key characterizations of ***h***TEGs for use in wearable electronics is the study of their long-term thermal stability and resistance to everyday wear. To this end, the evolution of the device's internal resistance was monitored over a period of 24 weeks under laboratory ambient conditions, with temperatures ranging between 19 and 22 °C and relative humidity between 58 and 66%. As shown in Fig. 7**a**, the internal resistance of the ***h***TEG gradually increased by approximately 20% of its initial value ($R_{I,0}$), indicating degradation of the device when exposed to common real-life conditions. This could be extrapolated to mean that the maximum output power may decrease by around 20% after 6 months. However, the origin of this degradation is not entirely clear, as it may be influenced by the intrinsic degradation of the ***h***TE material itself or of the electrode (silver paste). To clarify this, we replaced the silver paste with fresh silver contacts, and the internal resistance of the device after 6 months was found to be only 3% more than its initial value, as shown in Fig. 7**b**

To evaluate the stability of the ***h***TEG under wearable-like mechanical stress, we studied the evolution of the device's internal resistance as a function of the number of bending cycles around a cylinder with a diameter of 6 cm, simulating the approximate diameter of a human wrist. As shown in Fig. 7**c**, the internal resistance of the ***h***TEG remained approximately stable during the first 1200 bending cycles. Beyond this point, the increase became more pronounced, reaching about a 28% increase after 3000 cycles compared to the initial value. Once again, we sought to investigate the origin of this degradation, as it is known that silver electrodes are brittle and the PMMA matrix embedding the $Sb_2Te_3$ and $Bi_2Te_3$ NPs is not a flexible polymeric matrix. After completing the 3000 bending cycles, we replaced the silver electrodes with fresh ones, and the internal resistance of the ***h***TEG was essentially restored, maintaining 99.4% of its original value, as shown in Fig. 7**d** These two experiments demonstrate that the primary cause of ***h***TEG degradation lies in the silver electrodes used for device fabrication, rather than in the ***h***TE materials developed in this work.

Finally, to underscore the significance of FEM-guided design in ***h***TEG development, a device was fabricated with the same configuration and legs of uniform width (7 mm) and length (10 mm). **Supplementary Figs. 7a** and **8a** illustrates the evolution of the internal resistance of this ***h***TEG. Two distinct differences compared with the geometrically optimized ***h***TEG are evident: (1) the internal resistance of this ***h***TEG is significantly higher (4.2 kΩ vs. 180 Ω), and (2) due to the intrinsic mismatch in the electrical conductivities of $Sb_2Te_3$ (p-type) and $Bi_2Te_3$ (n-type), abrupt changes in internal resistance occur at the transitions between p-type and n-type legs. Since these two TE materials exhibit significantly different resistivities, if the leg cross-sections are not geometrically optimized, the current density distribution becomes discontinuous at the p–n transitions. This leads to localized resistance jumps at the interfaces, which manifest as abrupt resistance changes in the measured curves. In the optimized device, by tailoring the cross-sectional areas of the p- and n-type legs, we effectively match their electrical resistances, thereby homogenizing the current flow and minimizing internal resistance, which eliminates such abrupt transitions. **Supplementary Figs. 7b** and **8b** presents the power output and voltage measurements of the non-optimized device. In this case, the maximum power output under a $\Delta T$ of 30 K is only 14 nW. This power output value is consistent with previous reports on composites composed of $Bi(Sb)_2Te_3$ nanoparticles embedded in insulating polymer matrices such as PVDF or silk fibroin, where the maximum power output reaches around 10 nW for a thermal gradient of 30 K [29,30]. The power output value of the non-optimized ***h***TEG is not directly comparable to the optimized one, as the active areas of the two devices differ. When considering the respective areas of both ***h***TEGs, the optimized ***h***TEG achieves a maximum ***PoD*** of 40.37 nW cm$^{-2}$, while the non-optimized ***h***TEG exhibits a ***PoD*** of just 1.38 nW cm$^{-2}$. The FEM-assisted geometric optimization of the TEG has resulted in a net 30-fold improved device-generated power density ***efficiency***. It is therefore clear that, when comparing the performance of ***h***TEG devices without prior geometric optimization - such as the non-optimized device presented in this study and those previously reported by Weng et al. [29]. and Park et al.[30]with the geometrically optimized device developed here, the importance of conducting preliminary modeling of the TEG becomes evident. Such modeling is demonstrated to be a powerful tool to guide the design of TEGs/***h***TEGs with maximized power output per unit area.

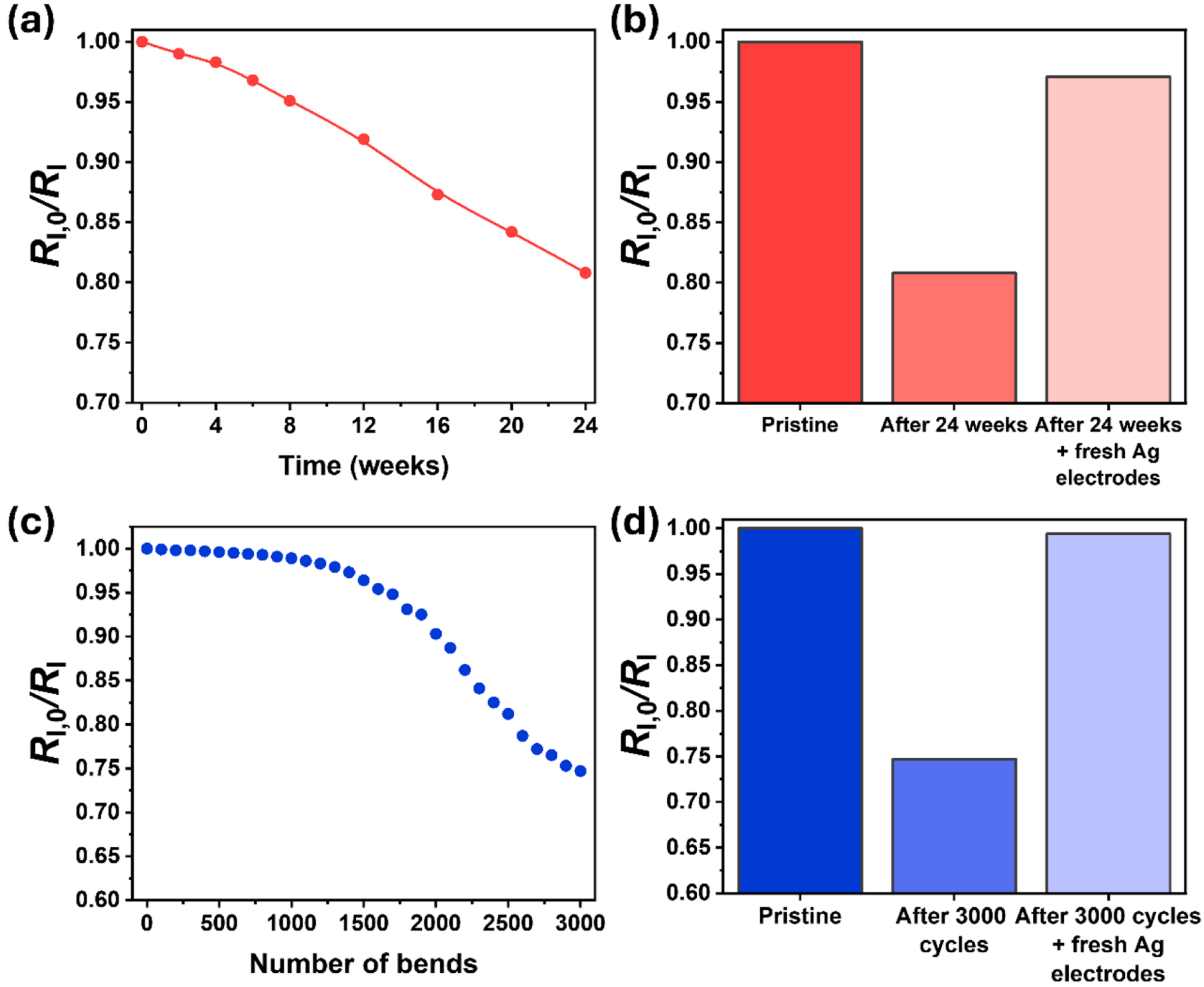

**Fig. 7.** (a) Long-term evolution of the ***h***TEG internal resistance ($R_I$). (b) Variation of internal resistance after 24 weeks and following the replacement of Ag electrodes. (c) Dependence of internal resistance ($R_I$) on the number of bending cycles. (d) Variation of internal resistance after 3000 bending cycles and subsequent replacement of Ag electrodes.

## 4. Conclusions

Nanostructured $Sb_2Te_3$ (p-type) and $Bi_2Te_3$ (n-type) with platelet morphology were successfully synthesized via a microwave-assisted thermolysis route and incorporated into hybrid thermoelectric (***h***TE) inks using PMMA as the polymer matrix. The use of an organothiol (HDT) in ink formulation has proven highly effective in enhancing the electrical conductivity of hybrid TE films by reducing the interface resistance. It is demonstrated through Hall mobility measurements that carrier mobility has increased significantly upon the addition of HDT into the hybrid-ink matrix. Through percolation studies, the optimal composition yielding the highest electronic power factor is determined to be 80% NP and 20% polymer content for both n-type ($Bi_2Te_3$) and p-type ($Sb_2Te_3$) hybrid TE films. A series of geometric and power optimizations has been performed using finite element modeling for a 6-leg ***h***TEG device, utilizing experimentally measured transport properties as materials' (individual leg's) characteristics. Results revealed a p-leg:n-leg area ratio of 1:2, with a p-leg width of 14 mm for the best power output from the fabricated hybrid films. A maximum power output of 1300 nW was estimated for the optimized TEG design, under a thermal gradient of 30 K. Thereafter, a TEG device with the identified parameters was fabricated, and its transport properties were evaluated under varying loads and temperature gradients. The highest power output of 950 nW was obtained at a 30 K thermal gradient, slightly lower than the theoretically predicted value, which is attributed to the contact resistance of the silver electrodes with the p- and n-legs, as demonstrated by TLM measurements. Furthermore, the optimized ***h***TEG achieves a maximum power output density (40.37 nW $cm^{-2}$) that is 30 times higher than that of the non-optimized ***h***TEG (1.38 nW $cm^{-2}$). This difference is attributed to the higher internal resistance of the non-optimized device. This work demonstrates that combining sustainable, low-energy chemical synthesis with simulation-guided device optimization enables the fabrication of flexible, high-performance ***h***TEGs. The proposed strategy not only minimizes energy use and material waste but also maximizes device efficiency, paving the way for more sustainable approaches to energy harvesting with TE materials.

## CRediT authorship contribution statement

**Bejan Hamawandi:** Writing – review & editing, Methodology, Data curation, Conceptualization. **José F. Serrano-Claumarchirant:** Writing – review & editing, Writing – original draft, Validation, Methodology, Formal analysis, Data curation, Conceptualization. **Adem B. Ergül:** Methodology, Investigation. **Kaspars Pudzs:** Investigation. **Inga Pudza:** Writing – review & editing, Visualization, Investigation, Data curation. **Parva Parsa:** Methodology, Investigation. **Metehan Kirsanli:** Investigation. **Oskars Bitmets:** Investigation. **Muhammet S. Toprak:** Writing

– review & editing, Writing – original draft, Supervision, Resources, Project administration, Investigation, Funding acquisition, Conceptualization.

## Declaration of competing interest

The authors declare that they have no known competing financial interests or personal relationships that could have appeared to influence the work reported in this paper.

## Acknowledgments

B.H. acknowledges the support of the Latvian Council of Science project No.1.1.1.9/LZP/1/24/043.

M.S.T. and J.F.S-C. acknowledges funding from the European Union's Horizon 2020 research and innovation program under grant agreement No 863222, and Olle Engkvist Foundation (SOEB, 226–0113). We acknowledge DESY (Hamburg, Germany), a member of the Helmholtz Association HGF, for providing experimental facilities. The experiments at the DESY PETRA III synchrotron were performed within proposal No I-20220381 EC. M.S.T. acknowledges the financial support from Olle Engkvist Foundation (SOEB, 190–0315) for establishing MW synthesis facilities. O.B and K.P. thank the support of the Latvian Council of Science, grant number lzp-2023/1–0456. J.F.S-C. acknowledges the support of the Juan de la Cierva fellowship (JDC2023-051223-I).

## Supplementary materials

Supplementary material associated with this article can be found, in the online version, at doi:10.1016/j.rineng.2026.110785.

## Data availability

Data will be made available on request.

## References

[1] G.J. Snyder, E.S. Toberer, Complex thermoelectric materials, Nat. Mater. 7 (2) (2008) 105–114, https://doi.org/10.1038/nmat2090.
[2] B. Hamawandi, H. Batili, M. Paul, S. Ballikaya, N.I. Kilic, R. Szukiewicz, M. Kuchowicz, M. Johnsson, M.S.M.-M. Toprak, High-efficiency nanostructured Bi2Te3 via high-throughput green solution chemical synthesis, Nanomaterials 11 (8) (2021) 2053, https://doi.org/10.3390/nano11082053.
[3] B. Hamawandi, S. Ballikaya, H. Batili, V. Roosmark, M. Orlovská, A. Yusuf, M. Johnsson, R. Szukiewicz, M. Kuchowicz, M.S. Toprak, Facile solution synthesis, processing and characterization of n- and p-type binary and ternary Bi–Sb tellurides, Appl. Sci. 10 (3) (2020) 1178, https://doi.org/10.3390/app10031178.
[4] M. Kober, T. Knobelspies, A. Rossello, L. Heber, Thermoelectric generators for automotive applications: holistic optimization and validation by a functional prototype, J. Electron. Mater. 49 (5) (2020) 2902–2909, https://doi.org/10.1007/s11664-020-07963-9.
[5] B. Singh Bhathal Singh, Thermoelectric generators: design, operation, and applications, in: New Materials and Devices for Thermoelectric Power Generation, 25, IntechOpen, 2023, pp. e275–e281, https://doi.org/10.5772/intechopen.1002722.
[6] M. Papapetrou, G. Kosmadakis, A. Cipollina, U. La Commare, G. Micale, Industrial waste heat: estimation of the technically available resource in the EU per Industrial sector, temperature level and country, Appl. Therm. Eng. 138 (2018) 207–216, https://doi.org/10.1016/j.applthermaleng.2018.04.043.
[7] R. Freer, A.V. Powell, Realising the potential of thermoelectric technology: a roadmap, J. Mater. Chem. C. Mater. 8 (2) (2020) 441–463, https://doi.org/10.1039/C9TC05710B.
[8] S. Yang, H. Ming, C. Zhu, Z. Wang, H. Xin, Z. Ge, D. Li, J. Zhang, X. Qin, High thermoelectric performance of N-type BiTeSe-based composites incorporated with both inorganic and organic nanoinclusions, ACS Appl. Mater. Interfaces. 16 (13) (2024) 16732–16743, https://doi.org/10.1021/acsami.4c02032.
[9] C. Ou, A.L. Sangle, A. Datta, Q. Jing, T. Busolo, T. Chalklen, V. Narayan, S. Kar-Narayan, Fully printed organic-inorganic nanocomposites for flexible thermoelectric applications, ACS Appl. Mater. Interfaces. 10 (23) (2018) 19580–19587, https://doi.org/10.1021/acsami.8b01456.
[10] Q. Xu, S. Qu, C. Ming, P. Qiu, Q. Yao, C. Zhu, T.R. Wei, J. He, X. Shi, L. Chen, Conformal organic-inorganic semiconductor composites for flexible thermoelectrics, Energy Env. Sci. 13 (2) (2020) 511–518, https://doi.org/10.1039/c9ee03776d.
[11] I. Paulraj, V. Lourdhusamy, C.J. Liu, Organic/inorganic n-type PVDF/Cu0.6Ni0.4 hybrid composites for thermoelectric application: a thermoelectric generator made of 8 pairs of p-leg ZnSb/Sb and n-leg β-PVDF/Cu0.6Ni0.4, Chem. Eng. J. 446 (2022) 137083, https://doi.org/10.1016/j.cej.2022.137083.
[12] Q. Jiang, J. Yang, P. Hing, H.R.A Ye, Design guidelines, and prospects of flexible organic/inorganic thermoelectric composites, Mater. Adv. 1 (5) (2020) 1038–1054, https://doi.org/10.1039/d0ma00278j.
[13] M. Massetti, F. Jiao, A.J. Ferguson, D. Zhao, K. Wijeratne, A. Würger, J. L. Blackburn, X. Crispin, S. Fabiano, Unconventional thermoelectric materials for energy harvesting and sensing applications, Chem. Rev. 121 (20) (2021) 12465–12547, https://doi.org/10.1021/acs.chemrev.1c00218.
[14] X. Zheng, Y. Hou, C. Bao, J. Yin, F. Yuan, Z. Huang, K. Song, J. Liu, J. Troughton, N. Gasparini, C. Zhou, Y. Lin, D.J. Xue, B. Chen, A.K. Johnston, N. Wei, M. N. Hedhili, M. Wei, A.Y. Alsalloum, P. Maity, B. Turedi, C. Yang, D. Baran, T. D. Anthopoulos, Y. Han, Z.H. Lu, O.F. Mohammed, F. Gao, E.H. Sargent, O.M. Bakr, Managing grains and interfaces via Ligand anchoring enables 22.3%-efficiency inverted perovskite solar cells, Nat. Energy 5 (2) (2020) 131–140, https://doi.org/10.1038/s41560-019-0538-4.
[15] W. Feng, Y. Tan, M. Yang, Y. Jiang, B.X. Lei, L. Wang, W.Q. Wu, Small amines bring big benefits to perovskite-based solar cells and light-emitting diodes, Chem. 8 (2) (2022) 351–383, https://doi.org/10.1016/j.chempr.2021.11.010.
[16] E.J.D. Klem, H. Shukla, S. Hinds, D.D. MacNeil, L. Levina, E.H. Sargent, Impact of dithiol treatment and air annealing on the conductivity, mobility, and hole density in PbS colloidal quantum dot solids, Appl. Phys. Lett. 92 (21) (2008) 1–4, https://doi.org/10.1063/1.2917800.
[17] J.F. Serrano-Claumarchirant, B. Hamawandi, A.B. Ergül, A. Cantarero, C. M. Gómez, P. Priyadarshi, N. Neophytou, M.S. Toprak, Thermoelectric inks and power factor tunability in hybrid films through all solution process, ACS Appl. Mater. Interfaces. 14 (17) (2022) 19295–19303, https://doi.org/10.1021/acsami.1c24392.
[18] H. Batili, B. Hamawandi, A. Björn Ergül, R. Szukiewicz, M. Kuchowicz, M. S. Toprak, A comparative study on the surface chemistry and electronic transport properties of Bi2Te3 synthesized through hydrothermal and thermolysis routes, Colloids Surf. Physicochem. Eng. Asp. 682 (2024) 132898, https://doi.org/10.1016/j.colsurfa.2023.132898.
[19] T.S. Sreeprasad, A.K. Samal, T. Pradeep, Bending and shell formation of tellurium nanowires induced by thiols, Chem. Mater. 21 (19) (2009) 4527–4540, https://doi.org/10.1021/cm901447z.
[20] S. Ferreira-Teixeira, A.M. Pereira, Geometrical optimization of a thermoelectric device: numerical simulations, Energy Convers. Manage 169 (2018) 217–227, https://doi.org/10.1016/j.enconman.2018.05.030.
[21] B. Şişik, S. LeBlanc, The influence of leg shape on thermoelectric performance under constant temperature and heat flux boundary conditions, Front. Mater. 7 (November) (2020) 1–13, https://doi.org/10.3389/fmats.2020.595955.
[22] Y. Thimont, S. Leblanc, The impact of thermoelectric leg geometries on thermal resistance and power output, J. Appl. Phys. 126 (9) (2019) 0–11, https://doi.org/10.1063/1.5115044.
[23] A. Rjafallah, D.T. Cotfas, P.A. Cotfas, Legs geometry influence on the performance of the thermoelectric module, Sustainability 14 (2022) 15823, https://doi.org/10.3390/su142315823.
[24] S. Skipidarov, M. Nikitin, in: S. Skipidarov, M. Nikitin (Eds.), Thin Film and Flexible Thermoelectric Generators, Devices and Sensors, Springer International Publishing: Cham, 2021, https://doi.org/10.1007/978-3-030-45862-1.
[25] X. Xiao, C.N. Kim, Y. Luo, X. Fan, Y. Deng, Performance analysis of a thermoelectric generator with a segmented leg, J. Electron. Mater. 48 (12) (2019) 7769–7779, https://doi.org/10.1007/s11664-019-07610-y.
[26] Wang, X.; Ting, D.S.-K.; Henshaw, P. The effects of geometry and substrate material on thermoelectric generator performance; 2020; pp 93–110. https://doi.org/10.1007/978-3-030-38804-1_6.
[27] B. Hamawandi, P. Parsa, I. Pudza, K. Pudzs, A. Kuzmin, S. Ballikaya, E. Welter, R. Szukiewicz, M. Kuchowicz, M.S. Toprak, Scalable solution chemical synthesis and comprehensive analysis of Bi2Te3 and Sb2Te3, Energy Mater. 5 (7) (2025) 500065, https://doi.org/10.20517/energymater.2024.204.
[28] Welter, E.; Chernikov, R.; Herrmann, M.; Nemausat, R. A beamline for bulk sample X-ray absorption spectroscopy at the high brilliance storage ring PETRA III; 2019; p 040002. https://doi.org/10.1063/1.5084603.
[29] Q. Guo, J. Guo, J. Zhou, W. Yu, P. Zhou, K. Yang, C. Zheng, S. Zeng, N. Hua, M. You, M. Weng, Patternable and flexible thermoelectric generators based on Bi2Te3/silk fibroin composites for temperature sensing and wearable applications, Chem. Eng. J. 500 (2024) 157077, https://doi.org/10.1016/j.cej.2024.157077.
[30] Y. Na, S. Kim, S.P.R. Mallem, S. Yi, K.T. Kim, K.-I.I. Park, Energy harvesting from Human body heat using highly flexible thermoelectric generator based on Bi2Te3 particles and polymer composite, J. Alloys. Compd. 924 (2022) 166575, https://doi.org/10.1016/j.jallcom.2022.166575.
[31] Y. Yao, Z. Chen, W. Wei, P. Zhang, Y. Zhu, Q. Zhao, K. Lv, X. Liu, Y. Gao, Cs0.32WO3/PMMA nanocomposite via in-situ polymerization for energy saving windows, Sol. Energy Mater. Sol. Cells 215 (2020) 110656, https://doi.org/10.1016/j.solmat.2020.110656.
[32] I. Mejia, M. Estrada, M. Avila, Improved upper contacts PMMA on P3HT PTFTS using photolithographic processes, Microelectron. Reliab. 48 (11–12) (2008) 1795–1799, https://doi.org/10.1016/j.microrel.2008.08.005.
[33] H. Batili, B. Hamawandi, P. Parsa, A. Björn Ergül, R. Szukiewicz, M. Kuchowicz, M. Sadaka Toprak, Electrophoretic assembly and electronic transport properties of rapidly synthesized Sb2Te3 nanoparticles, Appl. Surf. Sci. 637 (2023) 157930, https://doi.org/10.1016/j.apsusc.2023.157930.

[34] J. Choi, B.H. Kim, Ligands of nanoparticles and their influence on the morphologies of nanoparticle-based films, Nanomaterials 14 (2024) 1685, https://doi.org/10.3390/nano14201685.

[35] P. Zhou, X. Qiao, D.C. Milan, S.J. Higgins, A. Vezzoli, R.J. Nichols, Enhanced charge transport across molecule-nanoparticle-molecule sandwiches † ‡, Phys. Chem. Chem. Phys. 7176 (2023) 7176, https://doi.org/10.17638/datacat.liverpool.ac.uk/1853.

[36] I. Kanelidis, T. Kraus, The role of ligands in coinage-metal nanoparticles for electronics, Beilstein. J. Nanotechnol. 8 (2017) 2625–2639, https://doi.org/10.3762/bjnano.8.263.

[37] B. Liu, J. Hu, J. Zhou, R. Yang, Thermoelectric transport in nanocomposites, Materials 10 (2017) 418, https://doi.org/10.3390/ma10040418.

[38] O. Bubnova, X. Crispin, Towards polymer-based organic thermoelectric generators, Energy Env. Sci. 5 (11) (2012) 9345, https://doi.org/10.1039/c2ee22777k.

[39] L. Zhang, H. Shang, Q. Zou, C. Feng, H. Gu, F. Ding, High-power-density and excellent-flexibility thermoelectric generator based on all-SWCNTs/PVP composites, Small 20 (27) (2024), https://doi.org/10.1002/smll.202306125.

[40] C. Kim, J.Y. Baek, D.H. Lopez, D.H. Kim, H. Kim, Interfacial energy band and phonon scattering effect in Bi2Te3-polypyrrole hybrid thermoelectric material, Appl. Phys. Lett. 113 (15) (2018), https://doi.org/10.1063/1.5050089.

[41] S. Maity, U.K.S. Parvin, S. Das, K. Chatterjee, Polymer chalcogenides—new smart materials for thermoelectric applications, Smart. Mater. Struct. 31 (7) (2022) 073001, https://doi.org/10.1088/1361-665X/ac7595.

[42] R. Nayak, P. Shetty, M. Selvakumar, B. Shivamurthy, A. Rao, K.V. Sriram, M. S. Murari, A. Kompa, Deepika Shanubhogue, U. Influence of microstructure and thermoelectric properties on the power density of multi-walled carbon nanotube/metal oxide hybrid flexible thermoelectric generators, Ceram. Int. 49 (23) (2023) 39307–39328, https://doi.org/10.1016/j.ceramint.2023.09.275.

[43] B. Xia, X.-L. Shi, L. Zhang, J. Luo, W.-Y. Chen, B. Hu, T. Cao, T. Wu, W.-D. Liu, Y. Yang, Q. Liu, Z.-G. Chen, Vertically designed high-performance and flexible thermoelectric generator based on optimized PEDOT:PSS/SWCNTs composite films, Chem. Eng. J. 486 (2024) 150305, https://doi.org/10.1016/j.cej.2024.150305.

[44] J. Geng, B. Wu, Y. Guo, C. Hou, Y. Li, H. Wang, Q. Zhang, High power factor N-type Ag 2 Se/SWCNTs hybrid film for flexible thermoelectric generator, J. Phys. D. Appl. Phys. 54 (43) (2021) 434004, https://doi.org/10.1088/1361-6463/ac10db.

[45] C.M. Kim, S. Kim, N.R. Alluri, B. Bae, M.A. Okirigiti, G.H. Kim, H.J. Park, H. Jang, C. Baek, M.-K. Lee, G.-J. Lee, K.-I. Park, Flexible hybrid thermoelectric films made of bismuth telluride-PEDOT:PSS composites enabled by freezing-thawing process and simple chemical treatment, Mater. Today Chem. 44 (2025) 102532, https://doi.org/10.1016/j.mtchem.2025.102532.

[46] R. Nayak, P. Shetty, S. M, A. Rao, S. K V, S. Wagle, S. Nayak, V. Kamath, N. Shetty, M. Saquib, Formulation and optimization of copper selenide/PANI hybrid screen printing ink for enhancing the power factor of flexible thermoelectric generator: a synergetic approach, Ceram. Int. 50 (14) (2024) 25779–25791, https://doi.org/10.1016/j.ceramint.2024.04.315.